\documentclass{article}

\usepackage{soul}
\usepackage{url}
\usepackage[hidelinks]{hyperref}
\usepackage[utf8]{inputenc}
\usepackage[small]{caption}
\usepackage{graphicx}
\usepackage{amsmath}
\usepackage{amsthm}
\usepackage{booktabs}
\usepackage{algorithm}
\usepackage{algorithmic}
\usepackage{natbib}

\usepackage{tikz}
\usetikzlibrary{arrows,positioning,fit,calc,shadows,backgrounds,intersections}
\usepackage{pgfplots}

\usepackage{booktabs}

\title{Bayesian Optimization in Materials Science: A Survey}

\author{
Lars Kotthoff \and Hud Wahab \and Patrick A. Johnson\\
Center for Artificially Intelligent Manufacturing\\
University of Wyoming}

\begin{document}

\maketitle

\begin{abstract}
Bayesian optimization is used in many areas of AI for the optimization of black-box processes and has achieved impressive improvements of the state of the art for a lot of applications. It intelligently explores large and complex design spaces while minimizing the number of evaluations of the expensive underlying process to be optimized. Materials science considers the problem of optimizing materials' properties given a large design space that defines how to synthesize or process them, with evaluations requiring expensive experiments or simulations -- a very similar setting. While Bayesian optimization is also a popular approach to tackle such problems, there is almost no overlap between the two communities that are investigating the same concepts. We present a survey of Bayesian optimization approaches in materials science to increase cross-fertilization and avoid duplication of work. We highlight common challenges and opportunities for joint research efforts.
\end{abstract}

\section{Introduction}

In many areas of AI, decades of research have resulted in many different
approaches for solving hard problems. Perhaps the best example is machine
learning, where dozens of different approaches just to classification are
available, with little guidance on what to use for a particular problem. While
random forests are usually a good choice, there are many problems where other
approaches provide better performance. Even human experts struggle to choose the
best approach for a given problem without significant experimentation though.
Worse, once a particular approach is chosen, its hyperparameters need to be set
for optimal performance -- or is a different approach actually better once its
hyperparameters have been optimized?

This issue is not unique to machine learning -- the optimization of the
hyperparameters of processes is a ubiquitous problem in AI and many other areas,
including materials science. While this can be done efficiently for some
applications, for example because gradients are available or other knowledge
about the optimized process can be leveraged, there are many scenarios where
nothing is known about the process -- it is a black box that can be evaluated,
but whose inner workings are not sufficiently understood to aid in the
optimization process. Such black boxes are much harder to optimize in practice.

Black-box processes are optimized by repeatedly evaluating them for different
hyperparameter settings and observing the effect changes in hyperparameter
values have -- much like scientists have studied natural phenomena for
millennia. In many cases, these evaluations are expensive and it is crucial to
minimize their number as the optimization would otherwise use too many
resources. Bayesian optimization (BO) is a methodology that allows for just
that. It is a sample-efficient optimization method that, in general, does not
require a large number of samples to obtain good results and is thus
particularly suitable for black-box functions that are extremely expensive to
evaluate, as found in many areas of AI, engineering, and materials science,
where an evaluation may entail synthesizing and testing a new material and
require specialized equipment and skilled operators.

At the heart of modern Bayesian optimization approaches are
machine-learning-induced surrogate models, that learn to emulate the black-box
process to be optimized based on a small number of evaluations. Such surrogate
models have a long tradition in materials science where they helped in the
development of new materials and designs long before the advent of machine
learning. Traditionally, models were based on first-principles understanding and
physical laws that were painstakingly developed by scientists rather than the
data-driven approaches we see today.

Bayesian optimization, like surrogate models, is not a new concept, nor did it
originate in AI -- its origins can be traced back decades to engineering
applications. It has gained increasing popularity in materials science in the
last decade, similar to popularity gains in AI in the same time period. However,
research in both fields proceeds almost entirely independently, with no
cross-fertilization taking place even though the problems solved are the same in
many cases. This survey aims to bridge this gap and increase awareness in the AI
community on relevant research and approaches being developed in materials
science. We provide an overview of relevant recent research and highlight
challenges in materials science that AI research can help with and open problems
that may benefit from joint efforts.

\section{Background}

In modern Bayesian optimization, a surrogate model is fit based on the results
of evaluating the black-box process at different points in the hyperparameter
space. Such an initial design could be random, or based on another design of
experiments approach. In most cases, this surrogate model is induced by a
machine learning algorithm; Gaussian Processes and random forests are popular
choices. An acquisition function then determines the next point in
hyperparameter space to evaluate the expensive black-box process at, based on
the predictions of the cheap-to-evaluate surrogate model and possibly
uncertainty quantifications of those predictions. This acquisition function
balances exploitation, i.e.\ evaluating the neighborhood of points known to
yield good performance, and exploration, i.e.\ evaluating points that are far
from regions of the hyperparameter space with known performance to avoid getting
stuck in local optima. Commonly-used acquisition functions include expected
improvement and maximum uncertainty, which always explores.

The expensive black-box function is then evaluated at the point proposed in
this way to obtain ground-truth data that is combined with the data from the
initial design. The updated data is used to obtain the surrogate model for the
next iteration of the optimization process. The incremental augmentation of the
data the surrogate model takes into account iteratively improves and refines it
for the areas of interest, yielding more accurate approximations of the
black-box function and better optimization results. The Bayesian optimization
process can be stopped at any time when the desired quality is achieved or
resources are exhausted. While this makes it convenient to deploy in practice,
there are no guarantees of convergence to a global optimum except if an infinite
number of evaluations of the black-box function are allowed. As we make no
assumptions about the optimization landscape, the quality of the result cannot
be guaranteed or even assessed except by comparison to the results other methods
or repeated BO runs achieve.

\begin{figure}[htb]
\begin{center}
    \includegraphics[width=\columnwidth]{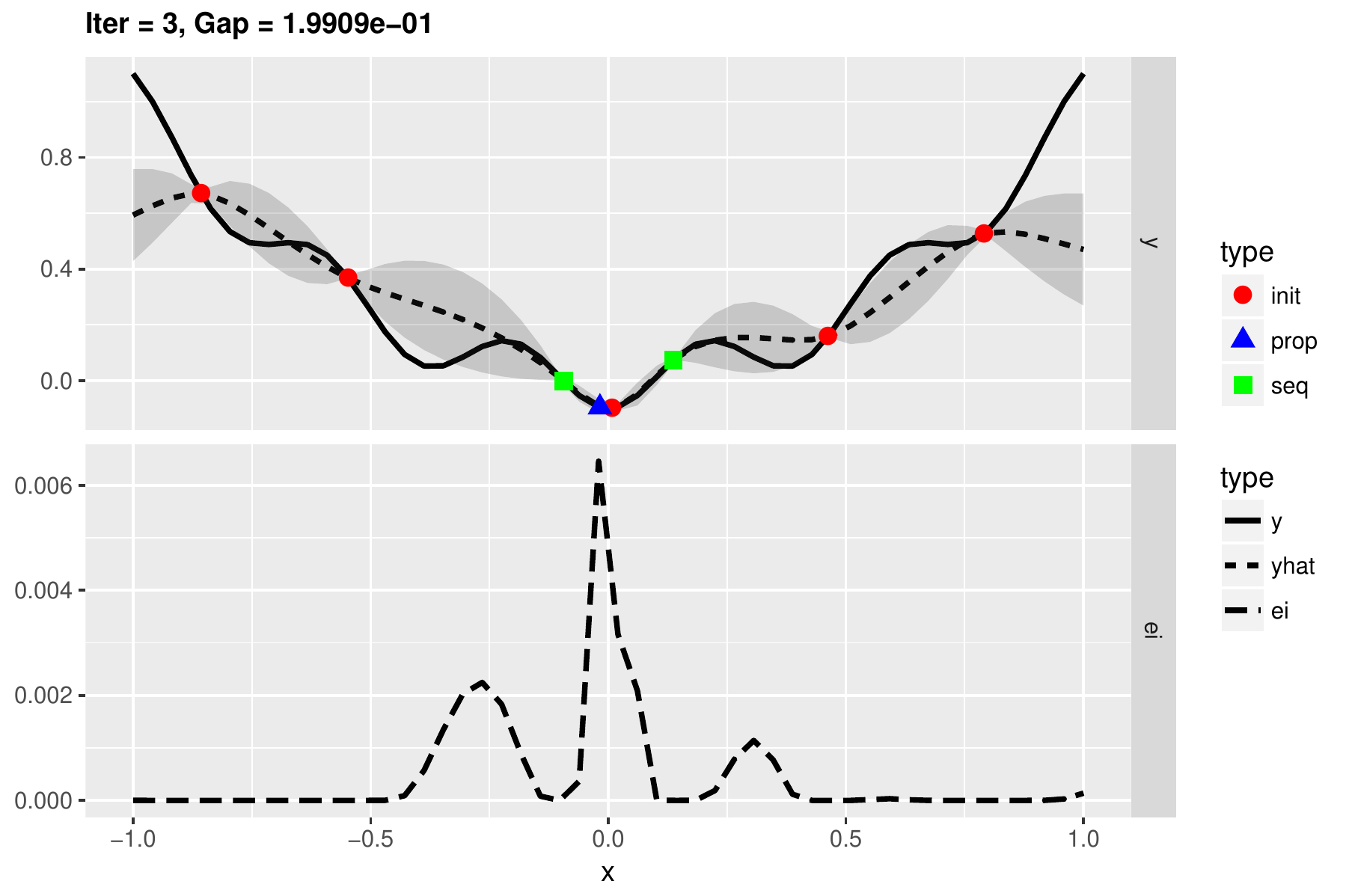}
    \caption{Example Bayesian optimization iteration. The horizontal axis shows
        the parameter space, projected into one dimension. In the top panel, the
        solid line is the true objective function; the dashed line is its
        surrogate-model-based approximation. The gray-shaded areas indicate the
        uncertainty of predictions of the surrogate model. Red circles represent
        evaluated configurations used to build the initial surrogate model;
        green squares configurations evaluated and added to the training data
        for the surrogate model in previous iterations; and the blue triangle
        the proposed next configuration to evaluate. The bottom panel shows the
        expected improvement over the best configuration found so far; the
        highest peak coincides with the proposed next configuration.
        Illustration generated with the \mbox{mlrMBO} package
        \protect\cite{bischl_mlrmbo_2017}.}
    \label{fig:mbo}
\end{center}
\end{figure}

Figure~\ref{fig:mbo} illustrates surrogate model and acquisition function for
one iteration of a toy example. There are many variations of this general
approach and hybrid methods that incorporate techniques from other areas of AI
have been developed, for example meta-learning to leverage known good
hyperparameter configurations from similar applications. A complete exposition
is beyond the scope of this survey. For more details on Bayesian optimization,
we refer the interested reader to
\cite{jones_efficient_1998,mockus_bayesian_1991} and
\cite{forrester_engineering_2008} for Bayesian optimization in engineering in
particular.

\section{Bayesian Optimization in AI}

The optimization of black-box processes is relevant in many areas of AI; in
particular for the automated hyperparameter tuning of algorithms. For many
applications, software to solve problems have hyperparameters that allow to
choose and tune, for example; a search heuristic, a stopping criterion, or a
pruning mechanism. These choices provide added flexibility and enable adaptation
of a generic algorithm to a particular problem, but setting them to achieve good
performance in a specific setting is a difficult task even for human experts.

There are numerous approaches that apply Bayesian optimization methods to
automated hyperparameter tuning in AI, for example
\cite{hutter_sequential_2011}, who introduce the SMAC system and demonstrate its
power tuning the hyperparameters of SAT and MIP solvers.
\cite{snoek_practical_2012} propose the spearmint system and automatically tune
the hyperparameters of a machine learning algorithm, whereas
\cite{feurer_efficient_2015,kotthoff_auto-weka_2017} automatically choose the
machine learning approach and pre- and post-processing methods in addition to
tuning its hyperparameters.

\cite{bischl_mlrmbo_2017} propose a general and flexible system for
hyperparamter tuning using Bayesian optimization, and \cite{falkner_bohb_2018}
combine Bayesian optimization with bandit-based methods. The need for
hyperparameter tuning in machine learning has resulted in the field of automated
machine learning (AutoML), which relies to a significant part on Bayesian
optimization and related methods. We refer the interested reader to a recent
book \cite{hutter_automated_2019} for more.

\section{Bayesian Optimization in Materials Science}

A large application area of Bayesian optimization in materials science is
similar to how it is applied in AI -- the optimization of the hyperparameters of
computationally expensive processes. In particular, Density Functional Theory
(DFT) simulations can compute properties of interest for a given material, but
may take weeks to complete. DFT calculations are used to determine the
electronic structure of systems composed of many atoms or molecules, in
particular the spatially dependent density of electrons, from which properties
of the modeled material can be derived. Such simulations rely only on
first-principles knowledge in quantum physics and are widely applicable to
many different types of materials. DFT and similar first-principles-based
simulations are the cornerstone in many areas of materials science. Note that
the DFT simulations themselves are not optimized, i.e.\ BO is not applied to
improve the simulation itself, but the underlying material -- DFT is used to
avoid having to synthesize and experimentally evaluate a material.

\cite{kiyohara_acceleration_2016,kikuchi2018,bondevik2019} use BO to optimize
the grain boundary structure in polycrystalline materials instead of
exhaustively evaluating the design space for the materials. They all demonstrate
that results of similar quality to exhaustive evaluation can be achieved at
significantly lower cost, with an increase in efficiency by up to two orders of
magnitude. Similarly, \cite{ueno_combo_2016} optimize the grain boundary energy
over several thousand precomputed hyperparameter configurations, which allows
them to compare the performance of different approaches.
\cite{talapatra_autonomous_2018} employ BO in a similar setting by evaluating
its performance on a set of a few hundred precomputed results. They optimize the
elastic properties of a material and demonstrate that BO is quickly able to
identify the optimal hyperparameters in their relatively small approximation of
the real search space. \cite{balachandran_adaptive_2016} investigate the same
application and in subsequent publications demonstrate the effectiveness of BO
to optimize shape memory alloys \cite{xue_accelerated_2016}, the band gap in
perovskites, an important material in the creation of solar cells
\cite{pilania_multi-fidelity_2017}, and the band gap in compounds for
luminescent materials \cite{lookman_active_2019}.
\cite{ling_high-dimensional_2017} demonstrate the utility of BO to optimize the
DFT-calculated magnetic deformation of a material, superconductors,
thermoelectricity, and the strength of steel. \cite{hankins_bio-like_2019} apply
BO to the problem of creating new materials via a parameterized generator for
patterns that describe the structure of the material. They optimize the
interfacial area of the material, which can be computed directly and very
cheaply compared to numerical simulations, but the large hyperparameter space
still necessitates a more efficient hyperparameter optimization approach than
exhaustive search. \cite{dehghannasiri_optimal_2017} design materials with low
energy dissipation by optimizing a dopant (an impurity introduced into a pure
material of a different type) and its concentration using BO, showing
improvements over random search and pure exploitation.

\cite{seko_machine_2014} perform one of the earliest investigations into whether
BO is suitable for optimizing the properties of materials that are evaluated
using DFT. In particular, they focus on the performance of the surrogate model
that replaces the DFT calculations, a crucial prerequisite for applying BO. They
predict and optimize the melting temperature of solid compounds and find that
support vector regression in particular provides high-quality predictions of the
DFT results. However, they use Gaussian Processes in a BO framework to optimize
the melting temperature, demonstrating better performance than random search. A
subset of the authors provide additional applications in a later study
\cite{tanaka_toward_2016} that more explicitly distinguishes between
high-throughput approaches where an exhaustive grid search is performed and
applications where this is infeasible and the use of BO or a similar technique
is necessary.

In contrast to most areas of AI, a lot of materials science is not only
computational, but involves physical experiments that synthesize and evaluate
materials. While DFT and similar simulations are increasingly accurate, their
results are often only approximations of what happens in the real world due to
different scales or a lack of first-principles understanding -- DFT calculations
can only consider relatively small numbers of atoms or molecules that may not
accurately represent materials deployed in practical applications. Even more so
than for computational simulations, BO is useful to reduce experimental effort
in the optimization of materials that requires not only physical resources, but
also in many cases specialized equipment and the time of a skilled operator.

\cite{ren_accelerated_2018} use a manual version of BO to discover metallic
glasses. They train a random forest model to predict whether a particular
composition of precursor materials results in the formation of a glass based on
prior experimental results. The model is then used to manually identify the
region with the highest promise for gathering additional data through
high-throughput experimentation, allowing for many evaluations in a short amount
of time. The experimental results obtained in this way are used to refine the
model. The authors iterate this process three times and discover several new
glass-forming systems. \cite{hase_phoenics_2018} use BO to optimize the
conditions for a chemical reaction to achieve a certain stability condition.
While the optimized experimental conditions could be evaluated through actual
experiments, the authors use a simulator to achieve higher throughput and obtain
more data. They show that BO is not only able to obtain the desired results
quickly, but also does so more robustly than other approaches, namely particle
swarm optimization and CMA-ES. \cite{kotthoff_ai_2019} optimize the reduction of
graphene oxide to graphene in the context of creating nano-circuits and flexible
electronics. The evaluations of the black-box function rely on physical
experiments that are performed manually and thus the number of data points is
only a few dozen. However, the authors demonstrate that even in this case,
significant improvements of the experimental outcome over human experts can be
achieved. \cite{wigley_fast_2016} optimize the production of Bose-Einstein
condensates with BO, which they refer to as machine learning online
optimization, with an order of magnitude fewer evaluations than previous
approaches. \cite{ren_embedding_2020} optimize the efficiency of solar cells by
tuning the growth temperature of gallium arsenide cells. Their customized BO
approach is able to achieve improvements after only five experimental
evaluations. \cite{vellanki2017} optimize alloy casting and the production of
polymer fibers in a constrained batch process, but do not provide a baseline
comparison for their results.

Some approaches combine computational simulations with experiments or different
types of simulations with different fidelities and costs for multiple layers of
screening. This allows to improve the outcomes of the optimization process while
reducing the costs for evaluations of the black box functions. Indeed, this is a
major difference to applications of BO in AI, where usually there is only a
single way of evaluating the black box function to be optimized. The closest to
evaluations at multiple levels are techniques like early stopping, used for
example in SMAC, that stop the evaluation if the hyperparameter setting under
configuration does not show promise on a subset of the data, e.g.\ a fraction of
the folds for a machine learning problem or a subset of the problem instances
for a combinatorial optimization algorithm.

\cite{gomez-bombarelli_design_2016} optimize the design of OLEDs. They
pre-screen designs using a machine learning surrogate model, then perform DFT
calculations on the remaining candidates, and finally synthesize and
experimentally evaluate the candidate designs that pass this second screening.
While the authors do not use BO directly, they demonstrate the power of a
multi-level, multi-fidelity approach in efficiently exploring a large search
space. \cite{pilania_multi-fidelity_2017} combine different types of DFT
simulations, one low-fidelity simulation that can be computed quickly, and one
high-fidelity simulation, to optimize the bandgap of solids. They use a modified
Gaussian Process that is able to take information at multiple fidelities into
account directly as a surrogate model and demonstrate better outcomes than when
using only the high-fidelity data. \cite{patra_2020} apply the same approach to
optimizing the bandgap of polymers and demonstrate similar results, noting
that their multi-fidelity surrogate model is able to generalize better to a
larger design space than surrogate models that use only high-fidelity data.

The majority of approaches in both AI and materials science consider
single-objective BO, as multi-objective optimization increases complexity
considerably and multiple objectives can be combined into a single objective.
However, a few approaches apply multi-objective BO to avoid requiring the user
to specify how to weight different objectives and being able to choose the
trade-off after the optimization from points on the Pareto front.
\cite{talapatra_autonomous_2018} consider the two competing objectives of shear
and bulk modulus, which quantify the effect external forces have on the
material. The authors use expected hypervolume improvement as the acquisition
function and show that they can efficiently identify the points on the Pareto
front. \cite{solomou2018} consider the simultaneous optimization of up to three
properties of a particular type of alloy based on computational simulations.
They also use expected hypervolume improvement as the acquisition function and
demonstrate the effectiveness of their approach. \cite{Ragasa_2019} scale up to
10 objectives, but use a custom non-BO approach, though inspired by Bayesian
optimization, that iteratively refines the hyperparameter search space based on
found Pareto-optimal configurations.

In many applications, it may be desirable to batch evaluations of the black box
function to be optimized, i.e.\ have the BO process predict multiple
hyperparameter configurations at once. This can facilitate the parallel
evaluation of configurations to make optimal use of available resources. This is
common to both AI and materials science, though there are few approaches that do
it. \cite{hase_phoenics_2018} propose batches of configurations to evaluate by
optimizing the acquisition functions for different values of a parameter that
trades off exploration and exploitation. The authors demonstrate that this batch
evaluation improves overall performance, as the correct setting for this
parameter is unclear and varies over time. A unique issue in BO for materials
science that does not usually come up in an AI setting is that the first
hyperparameter configuration to evaluate may constrain subsequent
configurations. \cite{vellanki2017} give the examples of heat treatment of
alloys, where multiple samples can be processed at the same time in an oven but
at a fixed temperature, and the production of polymer fibers, where different
values for polymer flow and coagulant speed can be evaluated at the same time
but within a fixed geometry. They propose a nested BO approach that optimizes
the hyperparameters that are subject to constraints in an outer loop and, given
the optimized values, the unconstrained hyperparameters in an inner loop.

\medskip

In many applications of BO in materials science, standard BO approaches, usually
with Gaussian Processes as surrogate models and expected improvement as
acquisition function, are used. There are a few exceptions; for example
\cite{hase_phoenics_2018} use Bayesian Neural Networks as a surrogate model with
a custom acquisition function and \cite{pilania_multi-fidelity_2017,patra_2020}
use custom Gaussian Processes that can take information at multiple fidelities
into account, with an acquisition function that takes the cost difference of the
different fidelities into account. \cite{dehghannasiri_optimal_2017} use mean
objective cost of uncertainty as the acquisition function, which quantifies how
much worse an outcome is because of uncertainty. \cite{ren_embedding_2020}
replace the traditional BO approach with multiple layers of surrogate models of
Bayesian networks infused with background knowledge that constrains the
(intermediate) outputs to physically feasible ones, combined with a neural
network that serves as a surrogate for expensive numerical simulations.
\cite{vellanki2017} propose a nested approach that runs a series of BO processes
to propose a batch of hyperparameter configurations to evaluate.
\cite{talapatra_autonomous_2018} use Bayesian model averaging inside the BO
process to select the most suitable surrogate model at the same time as
optimizing the black-box process. \cite{ueno_combo_2016} augment BO with
automated hyperparameter tuning and a few changes to be able to sample and
optimize the surrogate model more efficiently with thousands of features.

Most publications compare their approaches only to a grid or random search, but
some study the effect variations of the BO process have.
\cite{ling_high-dimensional_2017} propose a BO framework that uses random
forests as surrogate models with three different acquisition functions and
demonstrate that using likelihood of improvement instead of expected improvement
performs better on their case studies. \cite{balachandran_adaptive_2016} on the
other hand compare the performance of different surrogate models with the same
acquisition function. They note that models that quantify the uncertainty of
their predictions show better performance than those that do not, and find that
support vector regression with an RBF kernel gives the best surrogate models,
similar to the findings of \cite{seko_machine_2014}.


\medskip

In most cases, BO approaches in materials science are developed without
awareness of similar efforts in AI. A notable exception is
\cite{hase_phoenics_2018}, who compare to SMAC \citep{hutter_sequential_2011}
and spearmint \citep{snoek_practical_2012}, as well as non-BO optimization
methods. They demonstrate that their proposed system Phoenics achieves better
performance, differing from the other BO approaches in a different surrogate
model (Bayesian Neural Networks) and an acquisition function based on kernel
densities.

\section{Research Directions}

The application of Bayesian optimization in materials science is guided by the
domain. In particular, this means that achieving performance improvements
is perhaps not as important as in AI, and there is more emphasis on
understanding the models and being able to incorporate additional constraints
that encode the physical knowledge governing the application, as many of the
approaches mentioned above do.

Explainable AI has emerged as an important research area in AI with the
increasing deployment of AI models that facilitate decision making. In
particular deep neural network models are difficult to understand, with
real-world ramifications on their performance. Almost all publications in
materials science incorporate some analysis of what surrogate models have
learned, often with respect to the importance of the features used to
characterize the optimized process (e.g.~\cite{kotthoff_ai_2019}). To the best of
our knowledge, these approaches are relatively straightforward and more
sophisticated techniques developed by the AI community, for example the SHAP
framework \cite{lundberg_unified_2017}, have not been applied yet.

Another aspect that is emphasized more in materials science than in AI is the
quantification of the uncertainties of a surrogate model, which
\cite{ling_high-dimensional_2017} specifically develop for random forest
surrogate models. This may be the reason that Gaussian Processes remain the most
popular surrogate model, although different authors have observed that support
vector machine models achieve better predictive accuracy. This is a potential
research direction for the AI community that would benefit materials science
directly.

\cite{lookman_perspective_2016} highlight the successes machine learning in
general has had in materials science, but conclude that more applications are
required to give the community guidelines on how to apply BO and machine
learning in the wider sense in the context of materials science. While, as
outlined above, a standard incarnation of BO with Gaussian Processes as
surrogate models and expected improvement as acquisition function is emerging as
a reasonable place to start, there is likely no single approach that will
perform best in all settings -- a fact that has been known in AI for decades and
leveraged in the emerging field of automated machine learning
\cite{hutter_automated_2019} for example. While some approaches in materials
science are already taking advantage of these powerful techniques, a more
ubiquitous integration of AutoML techniques into materials science tools would
likely increase the uptake of BO in this area.

\cite{de_pablo_new_2019} identify a lack of systematic methods for reporting the
performance of machine learning approaches and baselines to compare to as one of
the challenges in materials science. This is again an area where AI research can
help, as the same issues are relevant and have been investigated longer. The
authors note that ``[e]stablished approaches from the statistics and computer
science communities combined with new methods developed specifically for
materials data issues must be disseminated to the materials community[\ldots]''
-- a challenge for our community to reach out more and organize
interdisciplinary events. Similarly, they note that most research in machine
learning focuses on large datasets which are usually not available in materials
science. Especially since the advent of deep learning, such small data problems
have been somewhat neglected by the AI community and there are opportunities to
develop new approaches at the intersection of materials and AI.

\cite{kalidindi_2019} mentions similar concerns and in particular notes that a
significant issue in materials science is the fragmented storage and
availability of the relevant data. Similar issues have been tackled by the AI
community and resulted in standard data formats and repositories that may be
applicable in a materials context as well, for example the OpenML project
\cite{vanschoren_openml_2013}. The author further highlights the challenge of
integrating the governing physical laws into a BO process and proposes a
framework to tackle this. Similar efforts to include background knowledge in
machine learning and data mining have been undertaken in AI, for example in the
ICON project \cite{bessiere_data_2016}, but no joint efforts exist to the best
of our knowledge.

While most research proceeds independently in AI and materials science, there
are a few joint efforts, notably \cite{vellanki2017}, which propose constrained
batch optimization and evaluate their approach both for a machine learning and
materials science applications. This demonstrates that both fields can benefit
from advances and makes a strong case for increased collaboration between AI and
materials science researchers.

\section{Conclusion}

We have presented an overview of applications of Bayesian optimization in
materials science, where researchers want to optimize poorly-understood
black-box processes to achieve desired properties of materials. Similar to
applications of BO in AI, the focus is on minimizing the number of evaluations
of the expensive black-box function while maximizing the quality of the result,
and similar results have been achieved, firmly establishing Bayesian
optimization as a state-of-the-art method that should be in the toolkit of every
materials science researcher.

A lot of recent progress in materials science has been enabled by Bayesian
optimization. However, a lot of work remains to be done to improve BO in
practice and scale to more difficult problems, in particular ones with multiple
objectives and large numbers of features, and integrate data-driven
methodologies with first-principles knowledge.

Materials science presents some unique challenges to applying Bayesian
optimization that can stimulate research in AI and, ultimately, benefit
applications of BO in AI as well. Research to date has progressed almost
independently in both application domains, and we hope that this survey will
stimulate interdisciplinary efforts and joint projects that advance either and
both fields.


\bibliographystyle{plainnat}
\bibliography{main}

\end{document}